\documentstyle[graphicx,aps,multicol]{revtex}
\begin{document}
\draft
\title{Thermal properties and radiative strengths in $^{160,161,162}$Dy}
\author{M.~Guttormsen\footnote{Electronic address: magne.guttormsen@fys.uio.no}, 
A.~Bagheri, R.~Chankova, J.~Rekstad, and S.~Siem}
\address{Department of Physics, University of Oslo, N-0316 Oslo, Norway}

\author{ A.~Schiller}
\address{Lawrence Livermore National Laboratory, 
L-414, 7000 East Avenue, Livermore CA-94551, USA}

\author{A.~Voinov}
\address{Frank Laboratory of Neutron Physics, Joint Institute of Nuclear
Research, 141980 Dubna, Moscow reg., Russia}
\maketitle

\begin{abstract}
The level densities and radiative strength functions (RSFs) in $^{160,161}$Dy have been extracted using the ($^3$He,$\alpha \gamma$) and ($^3$He,$^3$He$^{\prime} \gamma$) reactions, respectively. The data are compared to previous measurements on  $^{161,162}$Dy. The energy distribution in the canonical ensemble is discussed with respect to the nucleon Cooper pair breaking process. The gross properties of the RSF are described by the giant electric dipole resonance. The RSF at low $\gamma$-ray energies is discussed with respect to temperature dependency. Resonance parameters of a soft dipole resonance at $E_{\gamma}\sim 3$ MeV are deduced.
\end{abstract}


\begin{multicols}{2}

\section{Introduction}

The well-deformed rare earth region appears to be ideal for studying statistical properties of nuclei as function of temperature. The single particle Nilsson scheme displays almost uniformly distributed single particle orbitals with both parities. However, the low-temperature thermal properties of these nuclei are only poorly known. The main reason for this is the lack of appropriate experimental methods.

The Oslo Cyclotron group has developed a method to extract first-generation
(primary) $\gamma$-ray spectra at various initial excitation energies. From such a set of primary spectra, nuclear level density and radiative strength function (RSF) can be extracted \cite{hend1,schi0}. These two functions reveal essential nuclear structure information such as pair correlations and thermal and electromagnetic properties. In the last couple of years, the Oslo group has demonstrated several fruitful applications of the method \cite{melb0,schi1,gutt1,gutt2,schi2,gutt3,melb1,voin1,siem1}.

The subject of this work is to perform a systematic and consistent analysis of the three $^{160,161,162}$Dy isotopes. Since the proton number ($Z=66$) and the nuclear deformation ($\beta \sim 0.26)$ are equal for these cases, we expect to find the same electromagnetic properties. Furthermore, the underlying uniform distribution of single particle Nilsson states should from a statistical point of view give similar level densities for $^{160}$Dy and $^{162}$Dy. The present data set also allows us to check the results using the ($^3$He,$\alpha \gamma$) and ($^3$He,$^3$He$^{\prime}$$\gamma$) reactions for one and the same residual nucleus.

In Sect.~II an outline of the experimental procedure is given. The thermal aspects of the level density and RSF are discussed in Sects.~III and IV, respectively. Finally, concluding remarks are given in Sect.~V.

\section{Experimental method}

The experiments were carried out with 45 MeV $^3$He-ions at the Oslo Cyclotron Laboratory. Particle-$\gamma$ coincidences for $^{160,161,162}$Dy were measured with the CACTUS multi-detector array. The charged ejectiles were detected with eight particle telescopes placed at an angle of 45$^{\circ}$ relative to the beam direction. An array of 28 NaI $\gamma$-ray detectors with a total efficiency of $\sim$15\% surrounded the target and particle detectors. The following five reactions were utilized: $^{161}$Dy($^3$He,$\alpha \gamma$)$^{160}$Dy, $^{161}$Dy($^3$He,$^3$He$^{\prime}  \gamma$)$^{161}$Dy, $^{162}$Dy($^3$He,$\alpha \gamma$)$^{161}$Dy, $^{162}$Dy($^3$He,$^3$He$^{\prime}  \gamma$)$^{162}$Dy, and $^{163}$Dy($^3$He,$\alpha \gamma$)$^{162}$Dy. The three latter reactions have been reported earlier~\cite{melb0,schi1,schi2}. The reaction spin windows are typically $I\sim 2-6 \hbar$. The self-supporting targets are enriched to $\sim 95$\% with thicknesses of $\sim 2$ mg/cm$^2$. The experiments were run with beam currents of $\sim 2$ nA for 1--2 weeks.

The experimental extraction procedure and the assumptions made are described in Refs.~\cite{hend1,schi0} and references therein. For each initial excitation energy $E$, determined from the ejectile energy, $\gamma$-ray spectra are recorded. These spectra are the basis for making the first generation (or primary) $\gamma$-ray matrix \cite{gutt0}, which is factorized according to the Brink-Axel hypothesis \cite{brink,axel} as 
\begin{equation}
P(E,E_{\gamma}) \propto  \rho (E -E_{\gamma}) {\mathcal{T}}  (E_{\gamma}).
\label{eq:axel}
\end{equation}
Here, $\rho$ is the level density and ${\mathcal{T}}$ is the radiative transmission coefficient. 

The $\rho$ and ${\mathcal{T}}$ functions can be determined by an iterative procedure \cite{schi0} through the adjustment of each data point of these two functions until a global $\chi^2$ minimum with the experimental $P(E,E_{\gamma})$ matrix is reached. It has been shown \cite{schi0} that if one solution for the multiplicative functions $\rho$ and ${\mathcal{T}} $ is known, one may construct an infinite number of other functions, which give identical fits to the $P$ matrix by
\begin{eqnarray}
\tilde{\rho}(E-E_\gamma)&=&A\exp[\alpha(E-E_\gamma)]\,\rho(E-E_\gamma),
\label{eq:array1}\\
\tilde{{\mathcal{T}}}(E_\gamma)&=&B\exp(\alpha E_\gamma){\mathcal{T}} (E_\gamma).
\label{eq:array2}
\end{eqnarray}
Consequently, neither the slope ($\alpha$) nor the absolute values of the two functions ($A$ and $B$) can be obtained through the fitting procedure.

\begin{figure}

\begin{figure}
\includegraphics[totalheight=21cm,angle=0,bb=0 80 350 730]{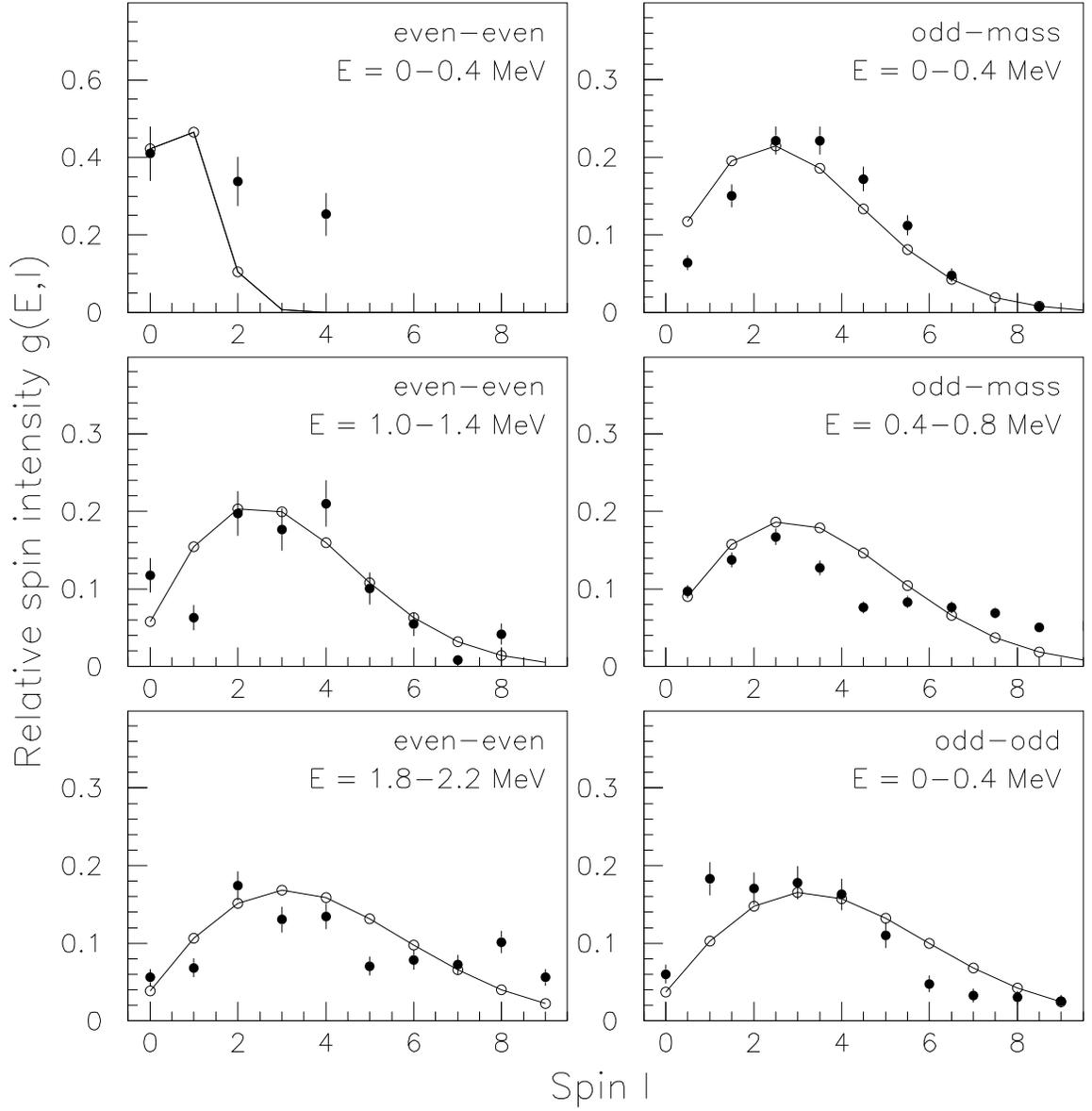}
\caption{Average experimental spin distributions (data points with error bars) compared to Eq.~(\ref{eq:spindis}). The data include 130 nuclei along the $\beta$--stability line in the $A=150-170$ mass region.}
\label{fig:spindis}
\end{figure}

\includegraphics[totalheight=21cm,angle=0,bb=0 80 350 730]{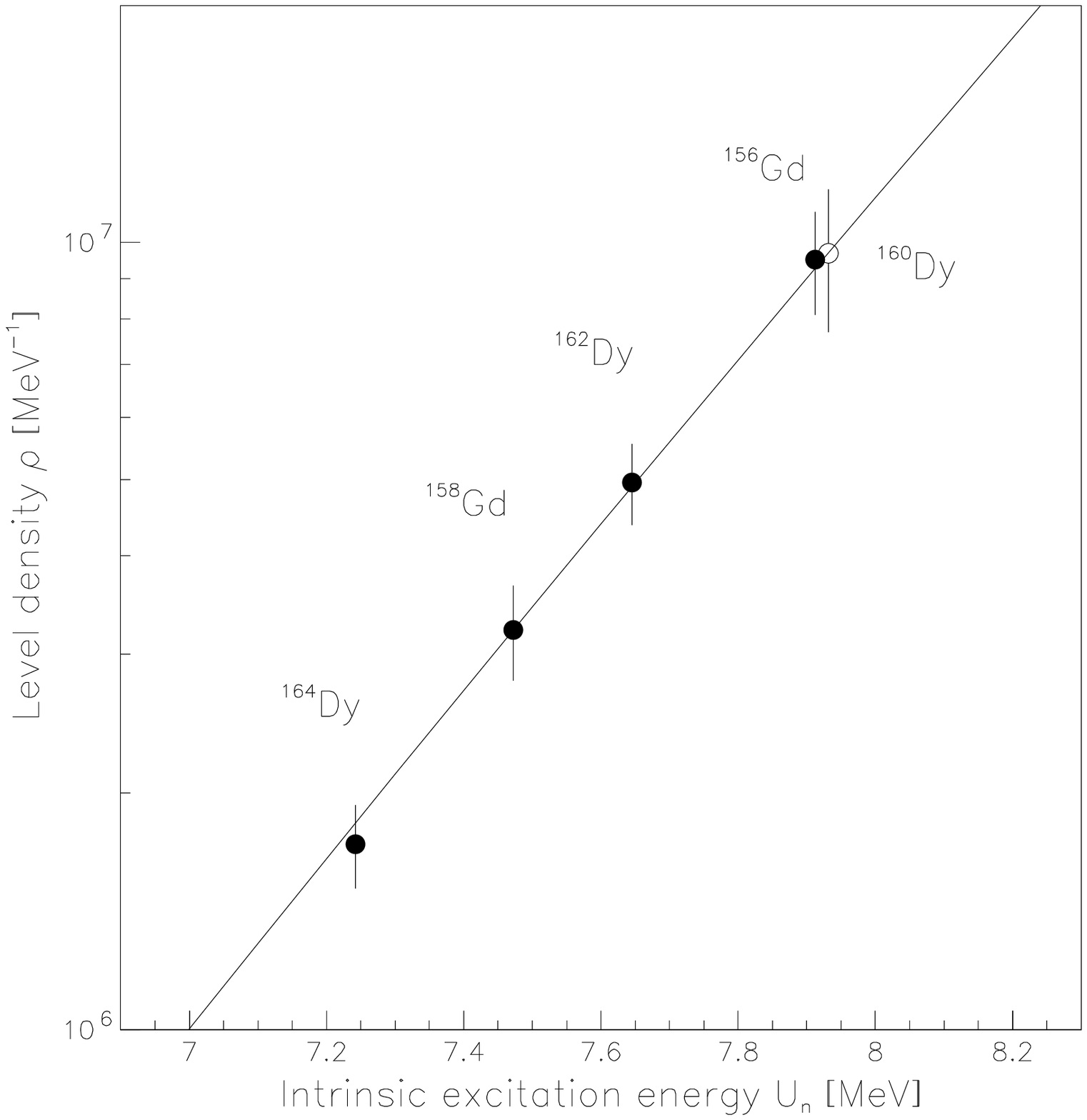}
\caption{Level densities estimated from neutron resonance level spacings at 
$B_n$. The data are plotted as function of intrinsic excitation energy $U_n=B_n-C_1-(\Delta_p+\Delta_n)$. The unknown level density for $^{160}$Dy (open circle) is estimated from the line determined by a least $\chi^2$ fit to the data points.}
\label{fig:rhosyst}
\end{figure}

The parameters $A$ and $\alpha$ can be determined by normalizing the level density to the number of known discrete levels at low excitation energy~\cite{ENSDF} and to the level density estimated from neutron-resonance spacing data at the neutron binding energy $E=B_n$~\cite{IA98}. The procedure for extracting the total level density $\rho$ from the resonance energy spacing $D$ is described in Ref.~\cite{schi0}. Since our experimental level density data points only reach up to an excitation energy of $E\sim B_n-1$ MeV, we extrapolate with the back-shifted Fermi gas model \cite{GC,egidy} 
\begin{equation} 
\rho_{\rm BS}(E)= \eta\frac{\exp(2 \sqrt{aU})}{12 \sqrt{2}a^{1/4}U^{5/4} \sigma_I}, 
\label{eq:bs}
\end{equation}
where a constant $\eta$ is introduced to fix $\rho_{\rm BS}$ to the experimental level density at $B_n$. The intrinsic excitation energy is estimated by $U=E-C_1-E_{\rm pair}$, where $C_1=-6.6A^{-0.32}$ MeV and $A$ are the back-shift parameter and mass number, respectively. The pairing energy $E_{\rm pair}$ is based on pairing gap parameters $\Delta_p$ and $\Delta_n$ evaluated from even-odd mass differences \cite{Wapstra} according to~\cite{BM}. The level density parameter is given by $a=0.21A^{0.87}$ MeV$^{-1}$. The spin-cutoff parameter $\sigma_I$ is given by $\sigma_I^2=0.0888aTA^{2/3}$, where the nuclear temperature is described by 
\begin{equation}
T=\sqrt{U/a}.
\label{eq:ua}
\end{equation}
In cases where the intrinsic excitation energy $U$ becomes negative, we put $U=0$, $T=0$ and $\sigma_I=1$. The spin distribution of levels (with equal energy) is given by \cite{GC}
\begin{equation}
g(E,I)=\frac{2I+1}{2\sigma_I ^2} \exp \left[-(I+1/2)^2 /2\sigma_I ^2\right],
\label{eq:spindis}
\end{equation}
which is normalized to $\sum_{I}g(E,I) \sim 1$. Figure~\ref{fig:spindis} compares $g(E,I)$ to the spin distributions of levels with known spin assignments~\cite{ENSDF}  for nuclei along the $\beta$-stability line with $A=150-170$. Although these data are incomplete and include systematical errors\footnote{One typical shortcoming of these compilations are that high spin members of rotational bands are overrepresented compared to low spin band heads.}, the agreement is gratifying and supports the expressions adopted for $\sigma_I$ and $g$.
 
\begin{figure}
\includegraphics[totalheight=21cm,angle=0,bb=0 80 350 730]{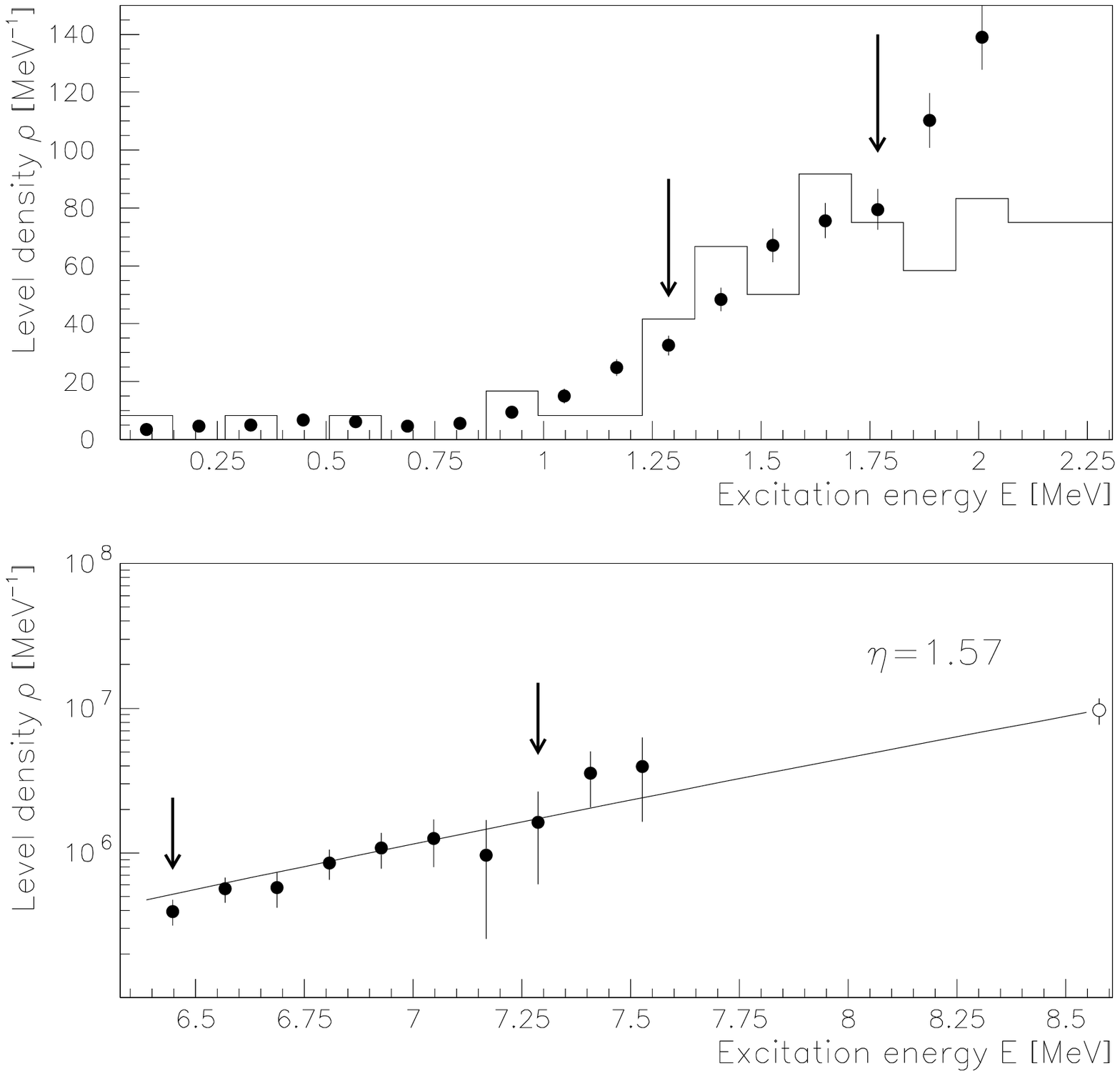}
\caption{Normalization procedure of the experimental level density (data points) of $^{160}$Dy. The data points between the arrows are normalized to known levels at low excitation energy (histograms) and to the level density at the neutron-separation energy (open circle) using the Fermi-gas level density (line).}
\label{fig:counting}
\end{figure}

Unfortunately, $^{159}$Dy is unstable and no information exists on the level density at $E=B_n$ for $^{160}$Dy. Therefore, we estimate the value from the systematics of other even-even dysprosium and gadolinium isotopes. In order to bring these data on the same footing, we plot the level densities as function of intrinsic energy $U$. From the systematics of Fig.~\ref{fig:rhosyst}, we estimate for $^{160}$Dy a level density of $\rho(B_n)=(9.7\pm 2.0) 10^{6}$ MeV$^{-1}$. Figure \ref{fig:counting} demonstrates the level density normalization procedure for the $^{160}$Dy case.
 
\begin{figure}
\includegraphics[totalheight=21cm,angle=0,bb=0 0 350 830]{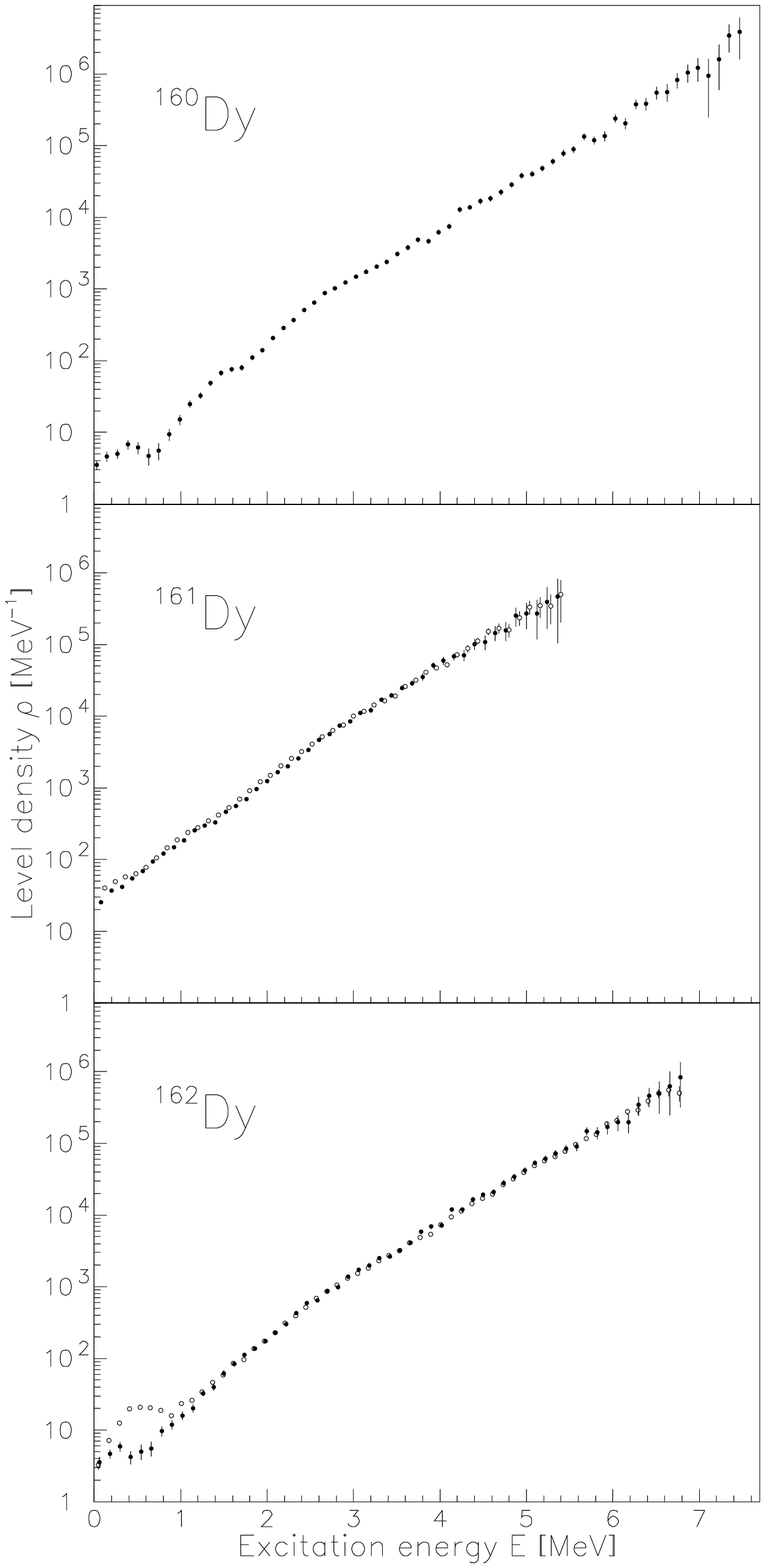}
\caption{Normalized level densities for $^{160,161,162}$Dy. The filled and open circles are measured with the ($^3$He,$\alpha$) and ($^3$He,$^3$He$^{\prime}$) reactions, respectively.}
\label{fig:rhoweb}
\end{figure}

The level densities extracted from the five reactions are displayed in Fig.~\ref{fig:rhoweb}. The data have been normalized as prescribed above, and the parameters used for $^{160,161,162}$Dy in Eq.~(\ref{eq:bs}) are listed in Table~\ref{tab:tab1}. The level densities for the three reactions previously published \cite{melb0,schi1,schi2} deviate slightly since we here have used updated and newly recommended data \cite{ENSDF,IA98}. The results obtained with the very different reactions ($^3$He,$\alpha$) and ($^3$He,$^3$He$^{\prime}$), are almost identical, except for the level density of the ground state band in $^{162}$Dy. Here, the ($^3$He,$^3$He$^{\prime}$) reaction overestimates the level density, as has been discussed previously~\cite{schi1}.

\section{Level density and thermal properties}

The level densities of $^{160}$Dy and $^{162}$Dy are very similar, however, $^{161}$Dy reveals several times higher level densities. In a previous work~\cite{gutt2}, it was claimed that the entropy for the excited quasi-particles is approximately extensive. To investigate this assumption further, we express the entropy as 
\begin{equation}
S(E)=k_{\rm B}\ln \Omega(E),
\end{equation}
where Boltzmann's constant is set to unity ($k_B=1$). The multiplicity $\Omega$ is directly proportional to the level density by $\Omega(E)= \rho(E)/\rho_0$. The normalization denominator $\rho_0$ is adjusted to give $S=\ln \Omega \sim 0$ in the ground state bands of the even-even nuclei. Here, we assume temperatures close to zero, thus fulfilling the third law of thermodynamics. The same $\rho_0$ is used for the odd-mass neighboring nuclei.

\begin{figure}
\includegraphics[totalheight=21cm,angle=0,bb=0 0 350 730]{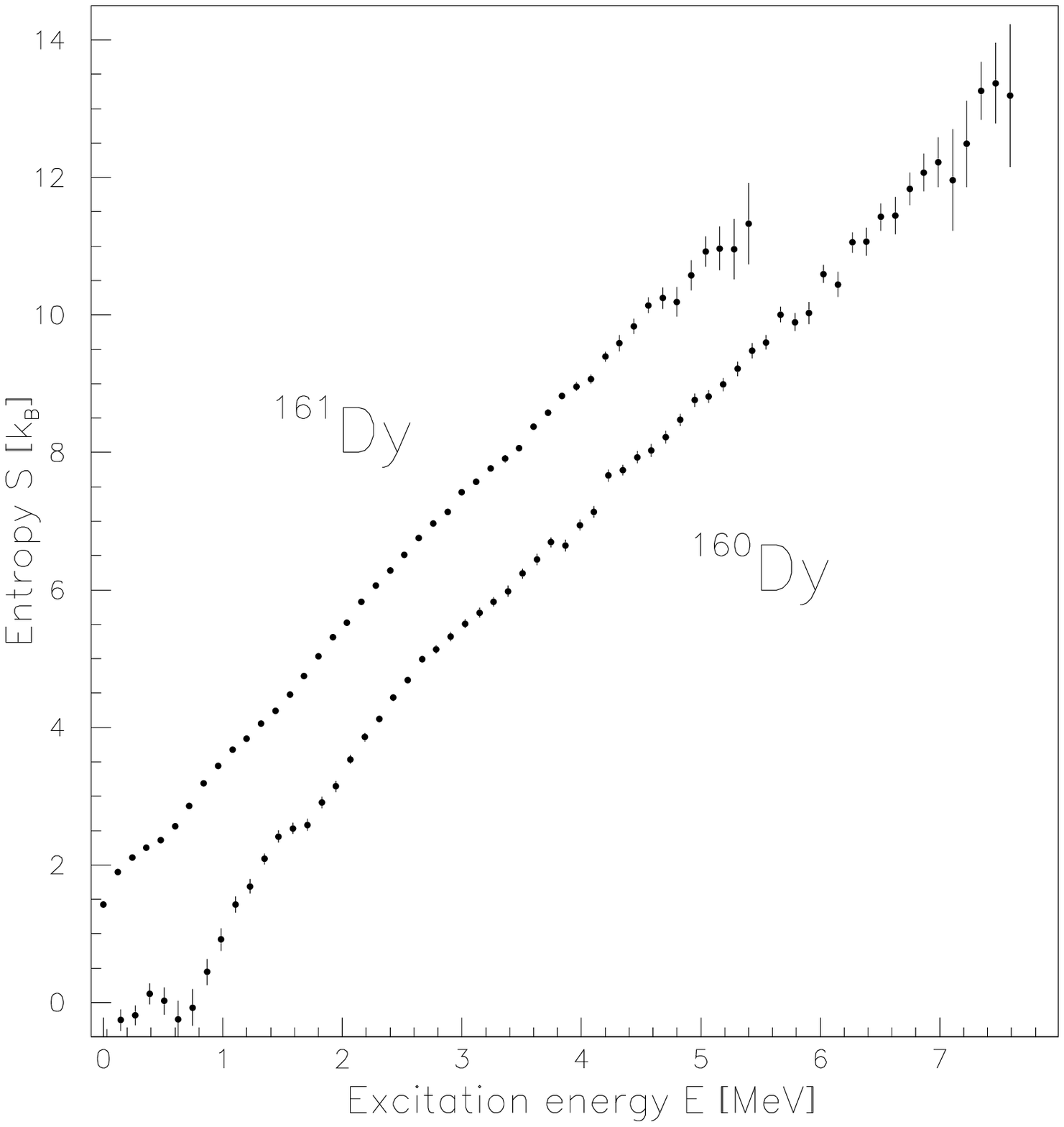}
\caption{Experimental entropy for $^{160,161}$Dy.}
\label{fig:dyentr}
\end{figure}

Figure~\ref{fig:dyentr} shows the entropies $S$ for the two new reactions reported in this work, i.e.,~the ($^3$He,$\alpha \gamma$)$^{160}$Dy and ($^3$He,$^3$He$^{\prime}\gamma$)$^{161}$Dy reactions. The results for the other reactions are very similar and are therefore not discussed here. The entropy of the $^{161}$Dy nucleus is seen to display an almost constant entropy excess compared to $^{160}$Dy. The difference, $\Delta S\sim2$, represents the entropy carried by the valence neutron outside the even-even $^{160}$Dy core (or hole coupled to the $^{162}$Dy core). It is an interesting feature that this difference is almost independent of excitation energy and therefore, of the number of quasi-particles excited in dysprosium, thus manifesting an entropy of $S_{qp}\sim2$ assigned to each quasi particle.

The probability that a system at fixed temperature $T$ has an excitation energy $E$, is described by the probability density function\footnote{The temperature $T$ is in units of MeV.}
\begin{equation}
p_T(E)=\frac{\Omega(E)\exp\left(-E/T\right)}{Z(T)},
\label{eq:pet}
\end{equation}
where the canonical partition function is given by
\begin{equation}
Z(T)= \sum_i\Delta\! E\;\Omega(E_i) \; e^{-E_i /T}.
\end{equation}
The experimental excitation energies $E_i$ have energy bins of $\Delta E$. In principle, the sum runs over all energies from zero to infinity, and we therefore use Eq.~(\ref{eq:bs}) to extrapolate to the higher energies. 
The energy distribution function $p_T(E)$ has a moment of the order $n$ about the {\em origin} given by
\begin{equation}
\langle E^n \rangle=\sum_i\Delta\! E\;E_i^n\;p_T(E_i).
\label{eq:eave}
\end{equation}

It is easy to show that the various moments also may be evaluated by the differentiation of $Z(T)$:
\begin{eqnarray}
\langle E  \rangle&=&\frac{T^2}{Z}\frac{dZ}{dT} \label{eq:zder1}\\
\langle E^2\rangle&=&\frac{T^4}{Z}\frac{d^2Z}{dT^2}+2T\langle E  \rangle \label{eq:zder2}\\
\langle E^3\rangle&=&\frac{T^6}{Z}\frac{d^3Z}{dT^3}+6T\langle E^2 \rangle -6T^2\langle E\rangle. \label{eq:zder3}
\end{eqnarray}

The moments $\mu_n$ of $E$ about its {\em mean} value $\langle E \rangle$ is defined by $\mu_n=\langle (E-\langle E \rangle )^n\rangle$. Thus, the second and third moments become
\begin{eqnarray}
\mu_2(T)&=&\langle E^2 \rangle-\langle E \rangle ^2
\label{eq:moments1}\\
\mu_3(T)&=&\langle E^3 \rangle-3\langle E^2 \rangle\langle E \rangle + 2\langle E \rangle ^3.
\label{eq:moments2}
\end{eqnarray}
These two moments are connected to the heat capacity and skewness of $p_T(E)$ according to 
\begin{eqnarray}
C_V&=&\mu_2/T^2 \\
\gamma &=&\mu_3/\mu_2^{3/2},
\end{eqnarray}
respectively. We also identify the standard deviation of the energy distribution as $\sigma_E=\sqrt{\mu _2}$.

\begin{figure}
\includegraphics[totalheight=21cm,angle=0,bb=0 0 350 830]{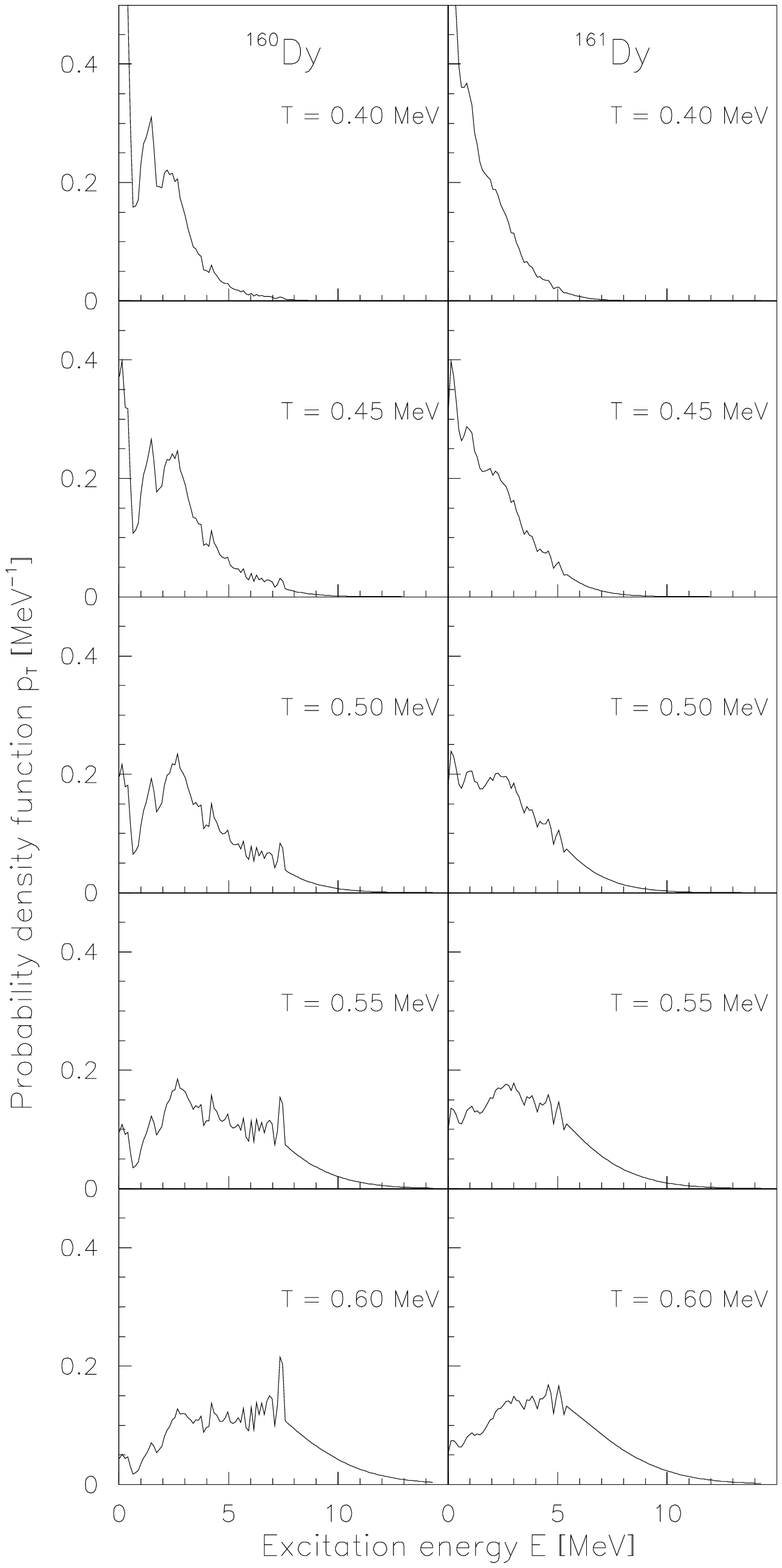}
\caption{Observed probability density functions for $^{160,161}$Dy.}
\label{fig:pet}
\end{figure}

Figure~\ref{fig:pet} shows the probability density functions for $^{160}$Dy and $^{161}$Dy. Below $T\sim 0.6$ MeV, the distribution is mainly based on experimental data, but at higher temperatures the influence of the somewhat arbitrary extrapolation of the level density by Eq.~(\ref{eq:bs}) will be increasingly important. The most interesting temperature region is around $T=0.5-0.6$ MeV, where the Cooper pair breaking process is at the strongest. At this point, the even-even and odd-even nuclei behave differently; $^{160}$Dy shows a broader distribution than $^{161}$Dy. This is due to the explosive behavior of $\rho$ for $E>E_{\rm pair}= 1.5-2$ MeV in even-even nuclei. Roughly, the number of levels for the breaking of neutron or proton pairs increases by a factor of $\exp(2S_{qp}) \sim 55$ giving totally $\sim 110$ times more levels. For the odd-even nuclei, Fig.~\ref{fig:rhoweb} shows that the level density is a monotonically increasing function from the ground state to higher excitation energies.

\begin{figure}
\includegraphics[totalheight=21cm,angle=0,bb=0 0 350 830]{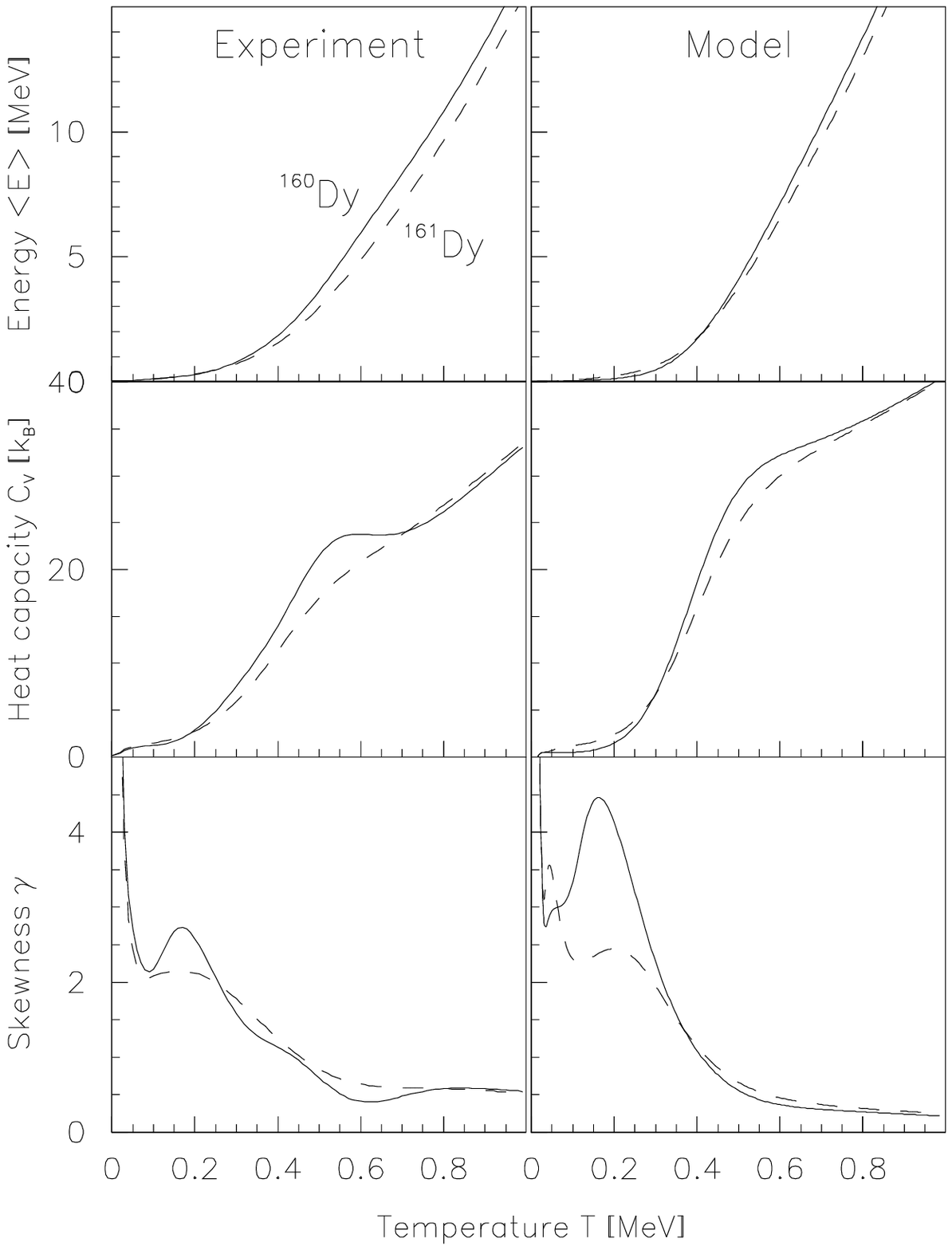}
\caption{Experimental (left) and theoretical (right) excitation energy $\langle E \rangle$, heat capacity $C_V$, and skewness $\gamma$ of the $p_T$ distribution as function of temperature $T$. The model parameters~\protect\cite{schi3} used are: $\varepsilon_p=\varepsilon_n=3a/\pi^2=0.19$~MeV, $\Delta_p=\Delta_n= 0.7$~MeV, $r=0.56$, $A_{\rm rigid}=7.6$~keV and $\hbar \omega_{\rm vib}=0.9$~MeV.}
\label{fig:tecg}
\end{figure}

The various experimental moments are best evaluated from $p_T(E)$, since the multiplicity $\Omega$ is directly known from the measured level densities. The left panels of Fig.~\ref{fig:tecg} show the corresponding values of average excitation energy $\langle E \rangle$, heat capacity $C_V$ and the skewness $\gamma$ of $p_T$ as function of temperature $T$. These key quantities characterize $p_T(E)$, and thereby reveal the thermodynamic properties of the systems studied. In the right panels these functions are compared to predictions evaluated in the canonical ensemble. The model~\cite{schi3} applied here treats the excitation of protons, neutrons, rotation, and vibration adiabatically with a multiplicative partition function 
\begin{equation}
Z=Z_{\pi} Z_{\nu} Z_{\rm rot} Z_{\rm vib},
\end{equation} 
where the various energy moments $\langle E^n \rangle$ are evaluated from Eqs.~(\ref{eq:zder1}-\ref{eq:zder3}).

The qualitative agreement between model and experiments shown in Fig.~\ref{fig:tecg} indicates that our model describes the essential thermodynamic properties of the heated systems. The heat capacity curves show clearly a local increase in the $T= 0.5-0.6$ MeV region, hinting the collective and massive breaking of nucleon Cooper pairs. This feature was recently discussed in Ref.~\cite{gutt4}, where two different critical temperatures were discovered in the microcanonical ensemble using the method of Lee and Kosterlitz \cite{lk90,lk91}: (i) The lowest critical temperature is due to the zero to two quasi-particle transition, and (ii) the second transition is due to the continuous melting of Cooper pairs at higher excitation energies. The first contribution is strongest for the even-even system ($^{160}$Dy), since the first broken pair represents a large and abrupt step in level density and thus a large contribution to the heat capacity. In $^{161}$Dy, the extra valence neutron washes out this step. The second contribution to $C_V$ is present in both nuclei signalizing the continuous melting of nucleon pairs at higher excitation energies. This second critical temperature appears at $\sim0.1$ MeV higher values.

The skewness $\gamma$ reveals higher order effects in the $p_T(E)$ distribution. For a symmetric energy distribution, $\gamma$ is zero. Figure~\ref{fig:tecg} shows positive values indicating distributions with high energy tails, as is confirmed by Fig.~\ref{fig:pet}. The $^{160}$Dy system shows a strong signal in $\gamma$ around $T\sim 0.2$ MeV. This signal is connected with the high energy tail of the $p_T(E)$ distribution into the $E > 2 \Delta$ excitation region with high level density.

\section{Radiative strength function and its resonances}

The slope of the experimental radiative transmission coefficient ${\mathcal{T}} (E_{\gamma})$ has been determined through the normalization of the level densities, as described in Sect.~II. However, it remains to determine $B$ of Eq.~(\ref{eq:array2}), giving the absolute normalization of ${\mathcal{T}}$. For this purpose we utilize experimental data~\cite{IA98} on the average total radiative width $\langle\Gamma_{\gamma} \rangle$ at $E=B_n$. 

We assume here that the $\gamma$-decay taking place in the continuum is dominated by E1 and M1 transitions and that the number of positive and negative parity states is equal. For initial spin $I$ and parity $\pi$ at $E=B_n$, the expression of the width~\cite{kopecky} reduces to
\begin{eqnarray}
\langle\Gamma_\gamma\rangle=\frac{1}{4 \pi \rho(B_n, I, \pi)} \sum_{I_f}&&\int_0^{B_n}{\mathrm{d}}E_{\gamma} B{\mathcal{T}} (E_{\gamma})
\nonumber\\
&&\rho(B_n-E_{\gamma}, I_f),
\label{eq:norm}
\end{eqnarray}
where the summation and integration run over all final levels with spin $I_f$ which are accessible by dipole ($L=1$) $\gamma$ radiation with energy $E_{\gamma}$. From this expression the normalization constant $B$ can be determined as described in Ref.~\cite{voin1}. However, some considerations have to be made before normalizing according to Eq.~(\ref{eq:norm}). 

\begin{figure}
\includegraphics[totalheight=21cm,angle=0,bb=0 80 350 730]{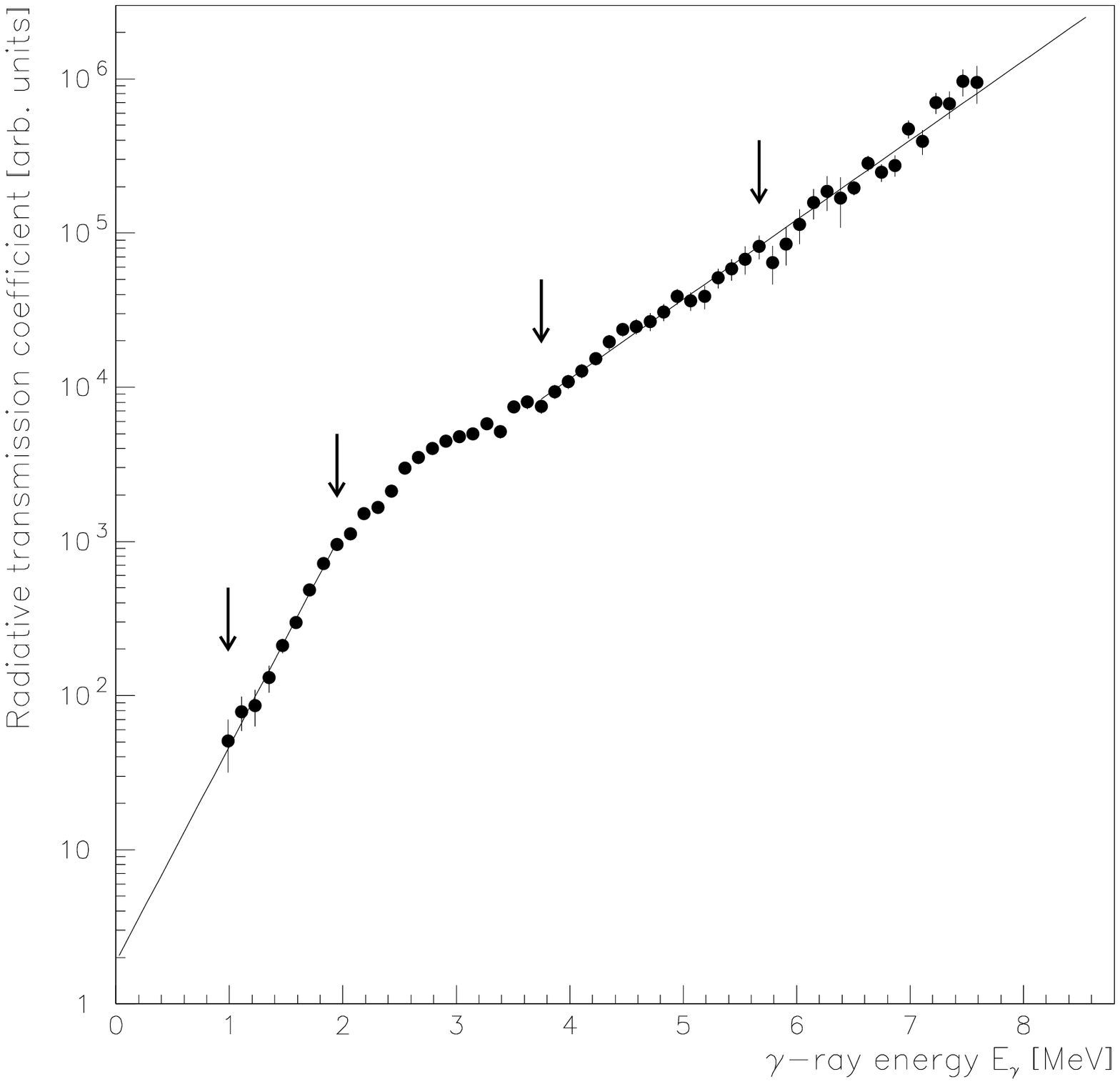}
\caption{Unormalized radiative transmission coefficient for $^{160}$Dy. The lines are extrapolations needed to calculate the normalization integral of Eq.~(\ref{eq:norm}). The arrows indicate the fitting regions.}
\label{fig:sigext}
\end{figure}

Methodical difficulties in the primary $\gamma$-ray extraction prevents determination of the functions ${\mathcal{T}}(E_{\gamma})$ and $\rho(E)$ in the interval $E_\gamma<1$~MeV and $E > B_n-1$~MeV, respectively. In addition, the data at the highest $\gamma$-energies, above $E_{\gamma} \sim B_n-1$~MeV, suffer from poor statistics. For the extrapolation of $\rho$ we apply the back-shifted Fermi gas level density of Eq.~(\ref{eq:bs}). For the extrapolation of ${\mathcal{T}}$ we use a pure exponential form, as demonstrated for $^{160}$Dy in Fig.~\ref{fig:sigext}. The contribution of the extrapolation to the total radiative width given by Eq.~(\ref{eq:norm}) does not exceed $15$\%, thus the errors due to a possibly poor extrapolation are expected to be of minor importance~\cite{voin1}.

For $^{160}$Dy, the average total radiative width at $B_n$ is unknown. However, the five $^{161-165}$Dy isotopes exhibit very similar experimental values of 108(10), 112(10), 112(20), 113(13), and 114(14) meV~\cite{IA98}, respectively. It is not clear how to extrapolate to $^{160}$Dy, but here the average value of $\langle\Gamma_{\gamma} \rangle =$ 112(20) meV has been adopted. 

\begin{figure}
\includegraphics[totalheight=21cm,angle=0,bb=0 10 350 830]{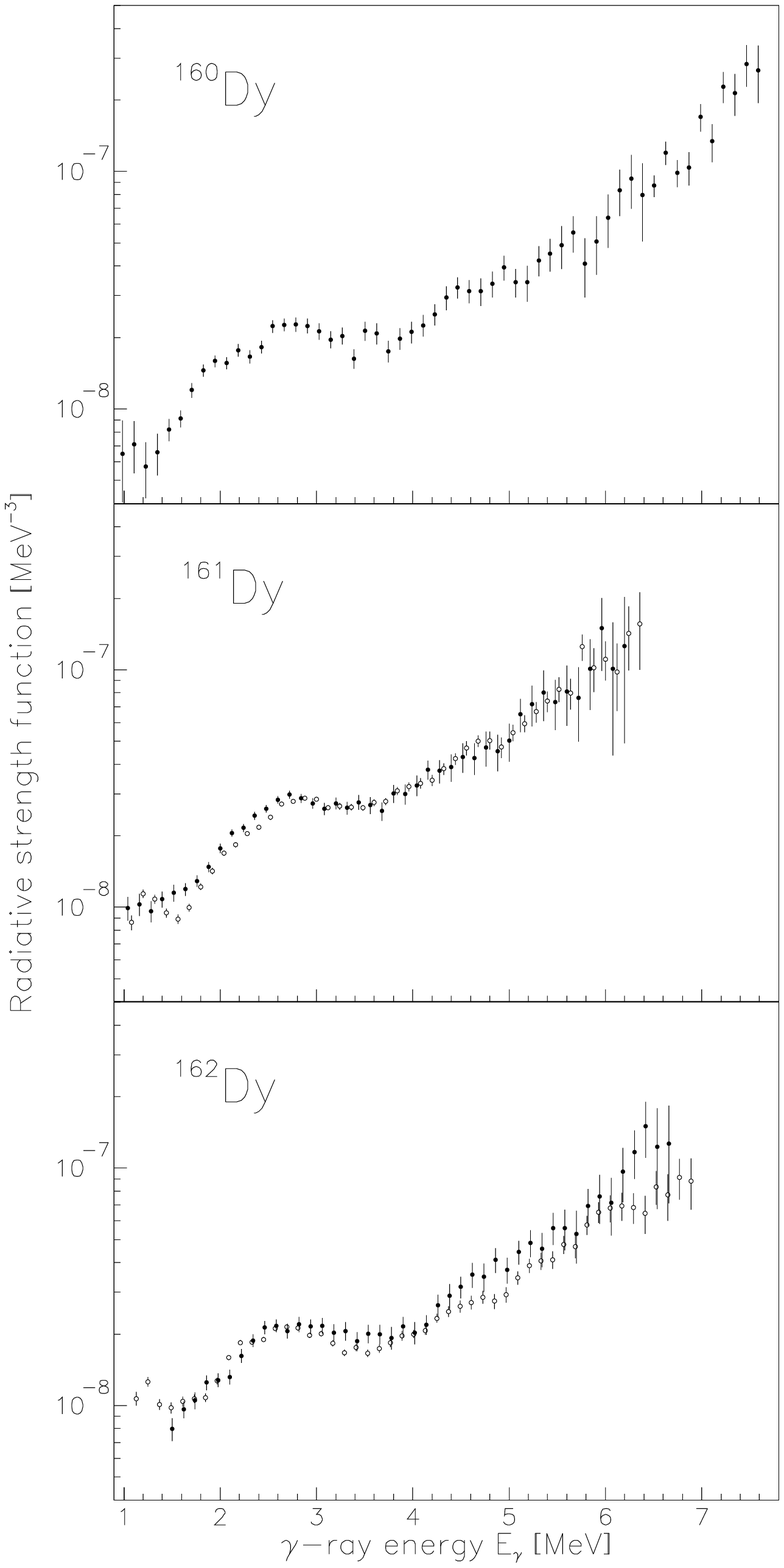}
\caption{Normalized RSFs for $^{160,161,162}$Dy. The filled and open circles are measured with the ($^3$He,$\alpha$) and ($^3$He,$^3$He$^{\prime}$) reactions, respectively.}
\label{fig:strfweb}
\end{figure}

The radiative strength function for $L=1$ transitions can be calculated from the normalized transmission coefficient by
\begin{equation}
f(E_{\gamma}) =\frac{1}{2\pi}\frac{ {\mathcal{T}} (E_{\gamma}) }{ E_{\gamma}^{3} }.
\end{equation}
The RSFs extracted from the five reactions are displayed in Fig.~\ref{fig:strfweb}. The data have been normalized with parameters from Tables~\ref{tab:tab1} and~\ref{tab:tab2}. Also here, the present results deviate slightly from the three data sets previously published \cite{melb0,schi1,voin1}. The present RSFs seem not to show any clear odd-even mass differences, and again the ($^3$He,$\alpha$) and ($^3$He,$^3$He$^{\prime}$) reactions reveal similar results.

The $\gamma$ decay probability is governed by the number and the character of available final states and by the RSF. A rough inspection of the experimental data of Fig.~\ref{fig:strfweb} indicates that the RSFs are increasing functions of $\gamma$-energy, generally following the tails of the giant electric (GEDR) and magnetic (GMDR) resonances in this region. In addition, a small resonance around $E_{\gamma}\sim 3$ MeV is found, the so-called pygmy resonance. These observations have been previously verified for several well deformed rare earth nuclei~\cite{melb0,voin1}. 

Also in the present work we adapt the Kadmenski{\u{\i}}, Markushev and Furman (KMF) model~\cite{kad} for the giant electric dipole resonance:
\begin{equation} 
f_{\mathrm{E1}}(E_\gamma)=\frac{1}{3\pi^2\hbar^2c^2} \frac{0.7\sigma_{\mathrm{E1}}\Gamma_{\mathrm{E1}}^2(E_\gamma^2+4\pi^2T^2)} {E_{\mathrm{E1}}(E_\gamma^2-E_{\mathrm{E1}}^2)^2}.
\label{eq:E1}
\end{equation}
Since the nuclei studied here have axially deformed shapes, the GEDR is split into two components GEDR1 and GEDR2. Thus, we add two RSFs with different resonance parameters,~i.e.,~strength $\sigma_{\rm E1}$, width $\Gamma_{\rm E1}$ and centroid $E_{\rm E1}$. The M1 radiation, which is supposed to be governed by the spin-flip M1 resonsance \cite{voin1}, is described by 
\begin{equation}
f_{\mathrm{M1}}(E_\gamma)=\frac{1}{3\pi^2\hbar^2c^2} \frac{\sigma_{\mathrm{M1}}E_\gamma\Gamma_{\mathrm{M1}}^2} {(E_\gamma^2-E_{\mathrm{M1}}^2)^2+E_\gamma^2\Gamma_{\mathrm{M1}}^2}.
\label{eq:M1}
\end{equation}
The GEDR and GMDR parameters are taken from the systematics of Ref.~\cite{IA98} and are listed in Table~\ref{tab:tab2}. The pygmy resonance is described with a similar Lorentzian function $f_{\rm py}$ as described in Eq.~(\ref{eq:M1}). Thus, we fit the total RSF given by
\begin{equation}
f=\kappa (f_{{\mathrm{E}}1} + f_{{\mathrm{M}}1})+f_{\mathrm{py}},
\end{equation}
to the experimental data using the pygmy-resonance parameters $\sigma_{\rm py}$, $\Gamma_{\rm py}$ and $E_{\rm py}$ and the normalization constant $\kappa$ as free parameters.

In previous works~\cite{melb0,voin1,siem1}, the temperature $T$ of Eq.~(\ref{eq:E1}) was also used as fitting parameter, assuming that a constant temperature could describe the data. The fitting to experimental data gave typically $T\sim 0.3$ MeV which is about the average of what is expected in this energy region. The use of a constant temperature approach is consistent with the Brink-Axel hypothesis~\cite{brink,axel}, which is utilized in order to separate $\rho$ and ${\mathcal{T}}$ through Eq.~(\ref{eq:axel}). 

\begin{figure}
\includegraphics[totalheight=21cm,angle=0,bb=0 0 350 730]{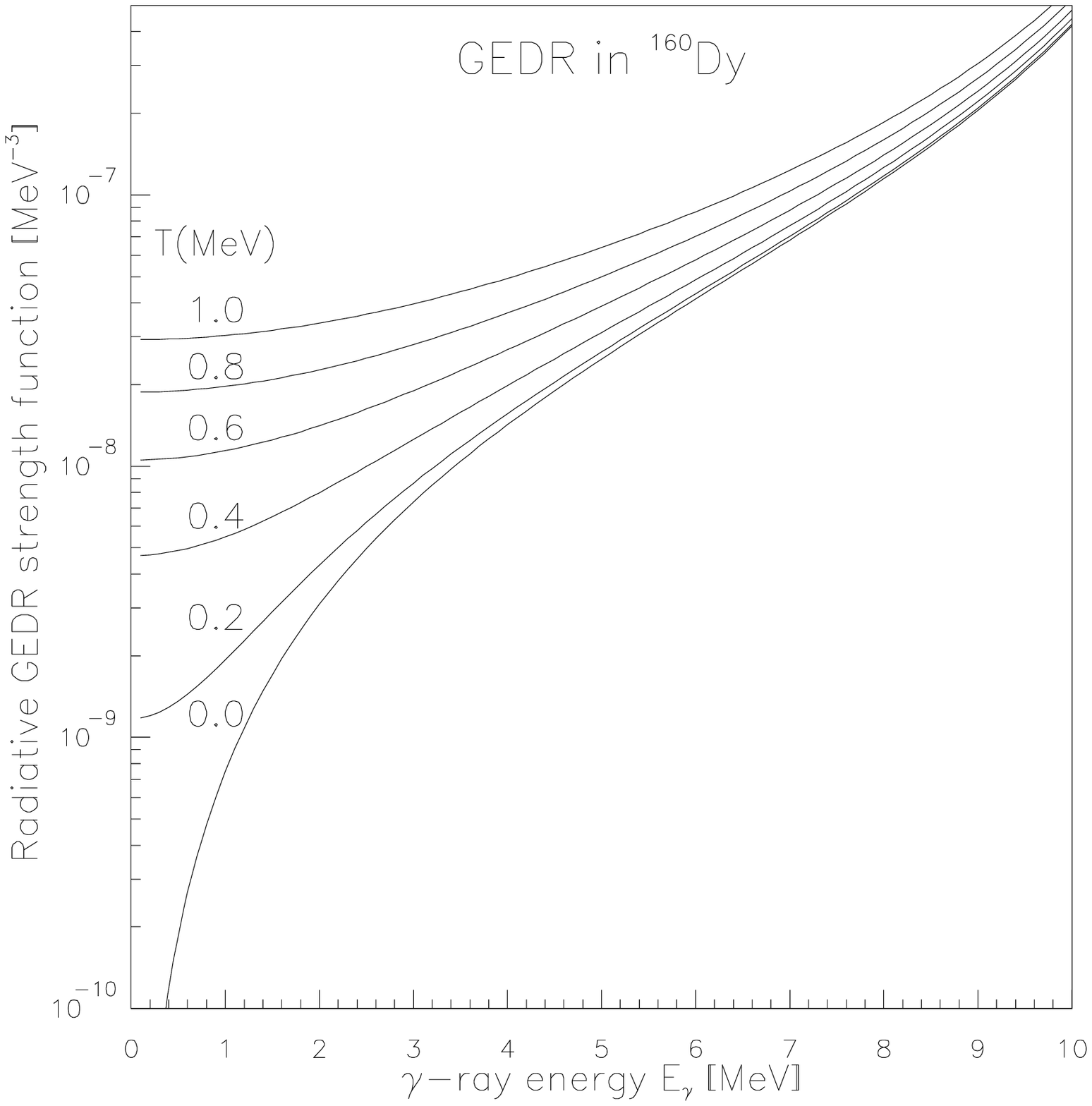}
\caption{Radiative GEDR strength function of the KMF model calculated for various temperatures.}
\label{fig:fteo}
\end{figure}

However, experimental data indicate that the RSF may depend also on how the temperature changes for the various final states. Data from the $^{147}$Sm(n,$ \gamma \alpha)^{144}$Nd reaction \cite{Popov} indicate a finite value of $f_{{\mathrm{E}}1}$ in the limit $E_\gamma\rightarrow 0$. Furthermore, in our study of the weakly deformed $^{148}$Sm, where no clear sign of the pygmy resonance is present, the RSF also flattens out at small $\gamma$ energies~\cite{siem1}. In the $^{56,57}$Fe isotopes it has been reported~\cite{emel1} that the RSF reveals an anomalous enhancement for $\gamma$ energies below 4 MeV. Also the $^{27,28}$Si isotopes show a similar increase in the RSF below 6 MeV~\cite{gutt5}. We should also mention that the extracted caloric curve $\langle E(T) \rangle$ of Fig.~\ref{fig:tecg} indicates a clear variation in $T$ for the excitation energy region investigated. Figure~\ref{fig:fteo} shows indeed that the strength of the tail of the GEDR, using the model of Eq.~(\ref{eq:E1}), is strongly temperature dependent. Therefore, from these considerations, we find it interesting to test the consequences by including a temperature dependent RSF in the description of the experimental data.

However, there is an inconsistency between such an approach and our extraction of the RSF using the Brink-Axel hypothesis through Eq.~(\ref{eq:axel}). The consequences have been tested in the following way: We first construct a typical level density and a temperature dependent transmission coefficient and multiply these two functions with each other to simulate a primary $\gamma$-ray matrix. Then Eq.~(\ref{eq:axel}) is utilized in order to extract $\rho$ and ${\mathcal{T}}$. It turns out that the output $\rho$ is almost identical with the input. Also ${\mathcal{T}}$ is correctly extracted, except small deviations of $\sim 15$\% for $\gamma$-energies below 1 MeV. Thus, the mentioned inconsistency should not cause severe problems.

\begin{figure}
\includegraphics[totalheight=21cm,angle=0,bb=0 0 350 730]{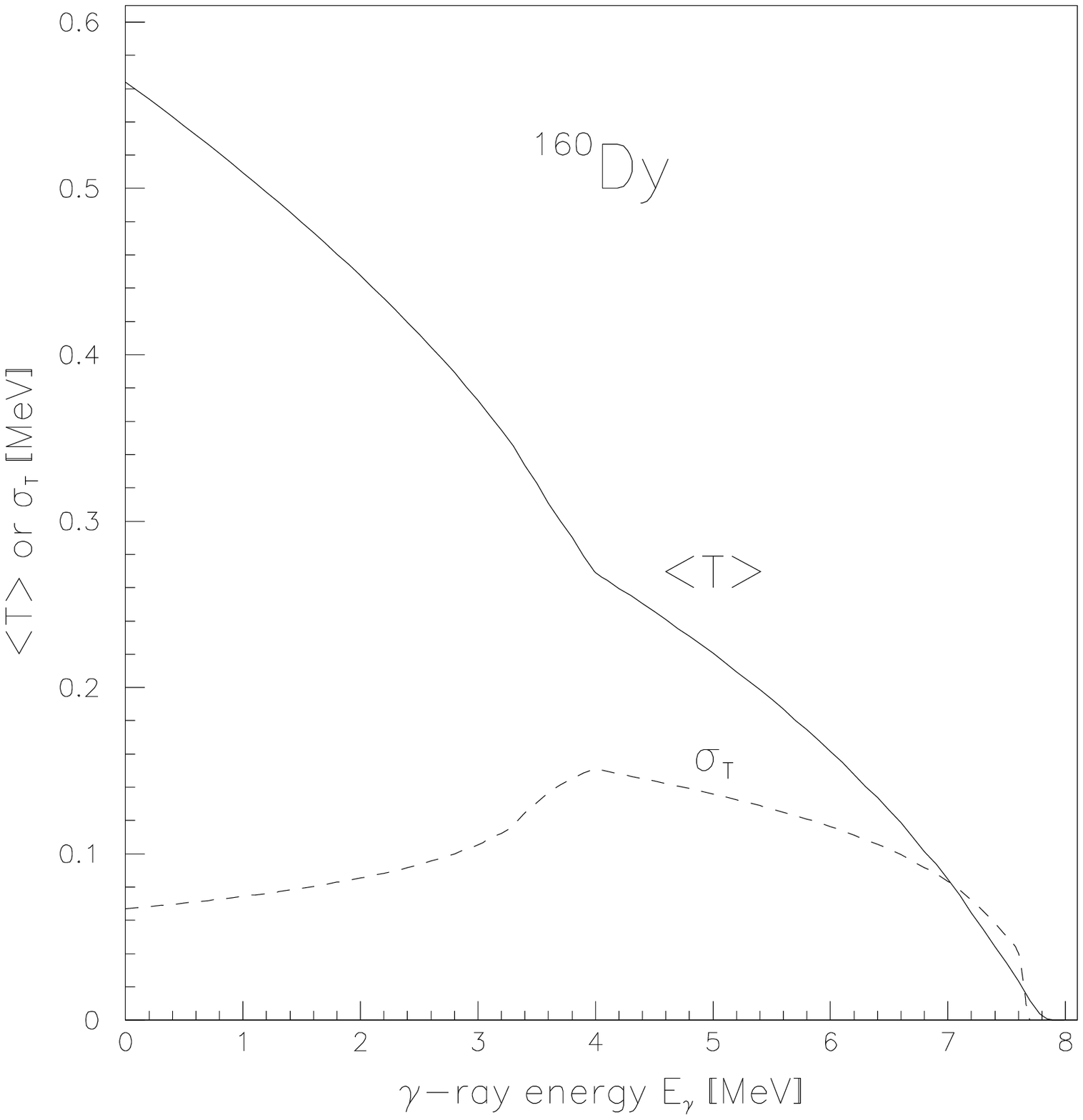}
\caption{Average temperature $\langle T\rangle$ of the final state (solid line) and standard deviation $\sigma_T$ (dashed line) for the temperature distribution as function of $\gamma$ energy in $^{160}$Dy, see text.}
\label{fig:tave}
\end{figure}

If we assume that the RSF depends on the temperature of the final states, it also depends on the primary $\gamma$-ray spectra chosen. Usually these spectra are taken at initial excitation energies between $E_1 \sim 4$ and $E_2 \sim 8$ MeV. Thus, the average temperature of the final states $E_f$ populated by a $\gamma$ transition with energy $E_{\gamma}$ is given by
\begin{equation}
\langle T(E_{\gamma}) \rangle=\frac{1}{E_2-E_1}\int_{E_1-E_{\gamma}}^{E_2-E_{\gamma}}{\mathrm{d}}E_f T(E_f),
\label{eq:Tave}
\end{equation}
where $T(E_f)=\sqrt{(E_f-C_1-E_{\rm pair})/a}$ is the schematic temperature dependency taken from Eq.~(\ref{eq:ua}). Figure~\ref{fig:tave} shows $\langle T \rangle$ and the standard deviation $\sigma_T=\sqrt{\langle T^2 \rangle -\langle T \rangle ^2 }$ for states populated with a $\gamma$ transition of energy $E_{\gamma}$ in $^{160}$Dy. The temperature goes almost linearly from 0.6 MeV to zero, giving an average of 0.3 MeV consistent with earlier constant temperature fits~\cite{melb0,voin1,siem1}. The standard deviation is relatively large, $\sigma_T\sim 0.1$ MeV, indicating that one should not replace $T$ by $\langle T \rangle$ in Eq.~(\ref{eq:E1}) but rather calculate $\langle f_{\mathrm{E1}}(E_\gamma) \rangle$ analog to the evaluation of $\langle T(E_{\gamma}) \rangle$ in Eq.~(\ref{eq:Tave}).

\begin{figure}
\includegraphics[totalheight=21cm,angle=0,bb=0 0 350 730]{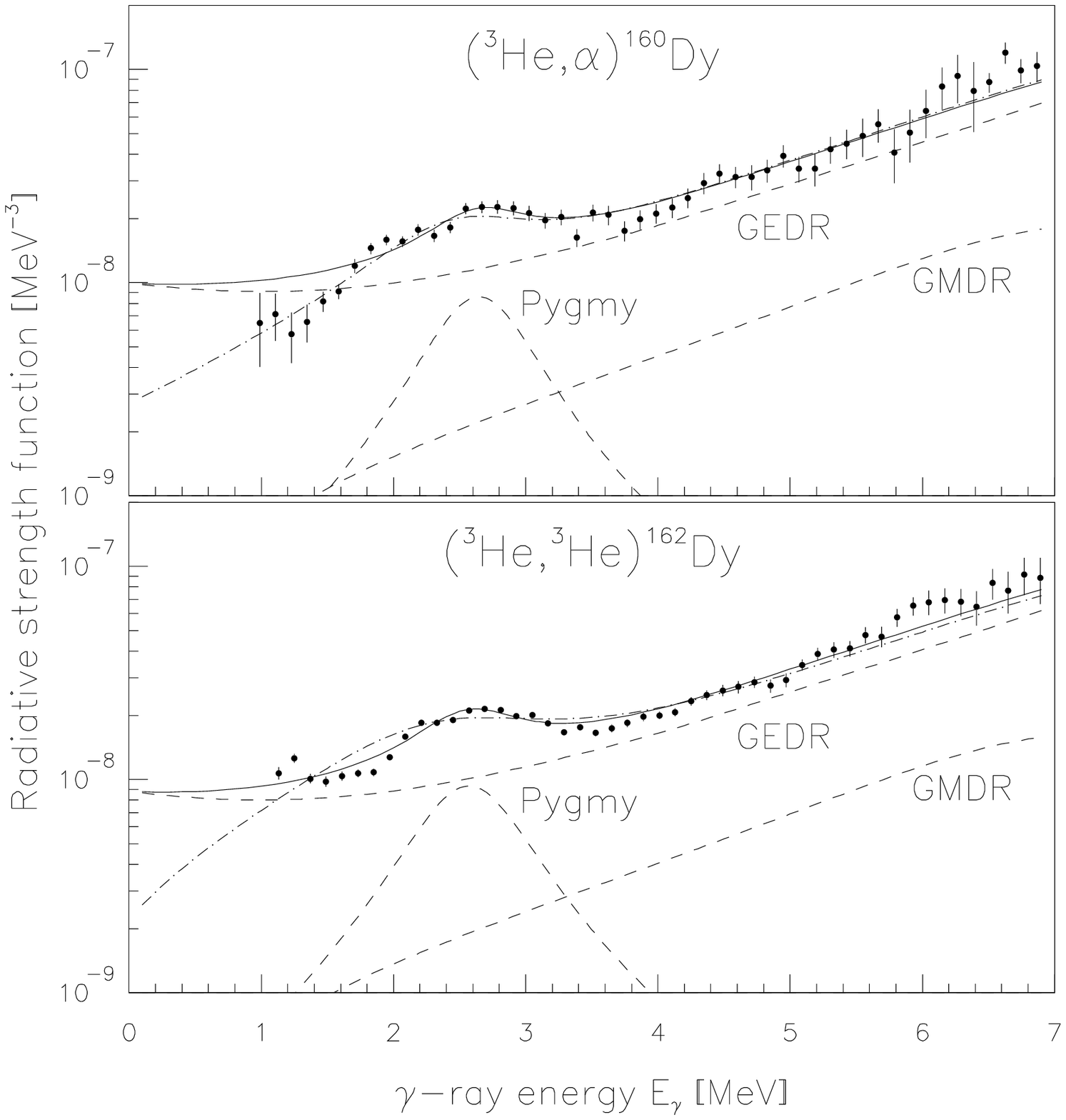}
\caption{The experimental RSFs for $^{160,162}$Dy (data points) compared to model predictions using a temperature dependent GEDR (solid line). The GEDR and pygmy resonance are the most important contributions (dashed lines) to the total RSF. The RSF using a fixed temperature of $T=0.3$ MeV (dash-dotted lines) gives significantly lower strength for $E_{\gamma}< 1$ MeV.}
\label{fig:fitexp}
\end{figure}

Figure \ref{fig:fitexp} shows fits to the experimental RSFs obtained from the ($^3$He,$\alpha$)$^{160}$Dy and ($^3$He,$^3$He$^{\prime}$)$^{162}$Dy reactions. The approaches using a varying temperature, $\langle f_{\mathrm{E1}}\rangle $, and a fixed temperature, $f_{\mathrm{E1}}(T=0.3$ MeV), are displayed as solid and dash-dotted lines, respectively. The $\langle f_{\mathrm{E1}}\rangle $ contribution to the total RSF is seen to give an increased strength for $E_{\gamma} < 1$ MeV, which the $^{162}$Dy case seems to support. However, the $^{160}$Dy case supports the fixed temperature approach. Unfortunately, the RSFs in the $E_{\gamma}\sim 1$ MeV region are experimentally difficult to measure. Here, a strong $\gamma$-decay intensity from vibrational states may not have been properly subtracted in the primary $\gamma$-ray spectra. Thus, the present data are not conclusive regarding the existence of enhanced radiative strength at low $\gamma$ energies.

In Table~\ref{tab:tab3}, we have summarized the fitted parameters for the pygmy resonance and the normalization constant $\kappa$ using varying temperatures. All the investigated dysprosium nuclei show similar pygmy resonance parameters and a mean average for all five reactions is given in the last row. Since the pygmy resonance is superimposed on the GEDR, the extraction of its resonance parameters is rather independent of how the GEDR temperature dependency is formulated.

\section{Summary and conclusions}

The present comparison between level densities and RSFs obtained with various reactions gives confidence to the Oslo method. The entropies of $^{161}$Dy follow very parallel the even-even $^{160,162}$Dy systems, assigning an entropy of $\sim 2$ to the valence neutron. The evolution of the probability density function with temperature was presented for $^{160,161}$Dy. The widths of these distributions increase anomalously in the $T=0.5-0.6$ MeV region. This feature of local increase in the canonical heat capacity is a fingerprint of the depairing process. Also the skewness of these distributions are studied showing the variation in the high energy tails as function of temperature. A simple canonical model is capable of describing qualitatively the various thermodynamic quantities.  

The five RSFs studied reveal very similar structures for all isotopes studied, as is expected since they all are considered to have the same deformation. The RSFs show a pygmy resonance superimposed on the tail of the giant dipole resonance. We have tested the consequences of introducing an RSF with varying temperatures in the GEDR case, which gives an enhanced strength at lower $\gamma$ energies. Our data are not conclusive in determining whether such effects are real or not.

\acknowledgements
Financial support from the Norwegian Research Council (NFR) is gratefully acknowledged. Part of this work was performed under the auspices of the U.S. Department of Energy by the University of California, Lawrence Livermore National Laboratory under Contract No.\ W-7405-ENG-48. A.V. acknowledges support from a NATO Science Fellowship under project number 150027/432 given by the Norwegian Research Council (NFR).

\end{multicols}

\newpage

\begin{table}
\caption{Parameters used for the back-shifted Fermi gas level density.} 
\begin{tabular}{l|ccc|ccc|c}
Nucleus    & $E_{\rm pair}$ & $a$         & $C_1$ &  $B_n$ & $D     $ & $\rho (B_n)$ &$\eta$ \\ 
           &     (MeV)      & (MeV$^{-1}$)& (MeV) & (MeV)  &   (eV)   & ($10^6$MeV$^{-1}$) & \\
\hline
&&&&&&\\

$^{160}$Dy &     1.945      & 17.37       & -1.301 & 8.576 &-       &9.7(20)$^a$ & 1.57     \\
$^{161}$Dy &     0.793      & 17.46       & -1.298 & 6.453 &27.0(50)&2.14(44) & 1.19     \\
$^{162}$Dy &     1.847      & 17.56       & -1.296 & 8.196 &2.4(2)&4.96(59) & 0.94     \\
\end{tabular}
$^a$Estimated from systematics.
\label{tab:tab1}
\end{table}

\begin{table}
\caption{Parameters used for the radiative strength functions.} 
\begin{tabular}{l|ccc|ccc|ccc|c}
Nucleus &$E^1_{{\mathrm{E}}1}$&$\sigma^1_{{\mathrm{E}}1}$&$\Gamma^1_{{\mathrm{E}}1}$&$E^2_{{\mathrm{E}}1}$&$\sigma^2_{{\mathrm{E}}1}$&$\Gamma^2_{{\mathrm{E}}1}$&$E_{{\mathrm{M}}1}$&$\sigma_{{\mathrm{M}}1}$&$\Gamma_{{\mathrm{M}}1}$&$\langle \Gamma_\gamma \rangle$\\
        &(MeV)     &(mb)           &(MeV)          &(MeV)     &(mb)           &(MeV)          &(MeV)   &(mb)         &(MeV)        &(meV) \\ \hline
&&&&&&&&&&\\
$^{160}$Dy& 12.47& 204.6& 3.22& 15.94& 204.6& 5.17&7.55& 1.51& 4.0& 112(20)$^a$ \\
$^{161}$Dy& 12.44& 206.0& 3.21& 15.92& 206.0& 5.14&7.54& 1.51& 4.0& 108(10) \\
$^{162}$Dy& 12.42& 207.5& 3.20& 15.90& 207.5& 5.12&7.52& 1.51& 4.0& 112(10) \\
\end{tabular}
$^a$Estimated from systematics.
\label{tab:tab2}
\end{table}

\begin{table}
\caption{Fitted pygmy resonance parameters and normalization constants.} 
\begin{tabular}{l|ccc|c}
Reaction&$E_{\mathrm{py}}$&$\sigma_{\mathrm{py}}$&$\Gamma_{\mathrm{py}}$& $\kappa$ \\
        &(MeV)            &(mb)                  & (MeV)                &   \\ \hline
&&&&\\
($^3$He,$\alpha$)$^{160}$Dy&2.68(25)  &0.27(11) &0.90(47)&1.06(12)  \\
($^3$He,$\alpha$)$^{161}$Dy&2.73(12)  &0.42(9) &0.95(24)&1.31(11) \\
($^3$He,$^3$He$^{\prime}$)$^{161}$Dy&2.86(7)   &0.40(4) &0.90(12)&1.27(5)  \\
($^3$He,$\alpha$)$^{162}$Dy&2.74(22)   &0.28(12) &0.78(34)&1.02(11)  \\
($^3$He,$^3$He$^{\prime}$)$^{162}$Dy&2.61(8)   &0.28(4) &0.98(18)&0.93(4)  \\ \hline
&&&&\\
Mean average                         &2.72(7)  &0.33(4) &0.90(13)&1.12(4) \\
\end{tabular}
\label{tab:tab3}
\end{table}

\clearpage

\end{document}